# Synthesis of ErFeAsO-based superconductors by hydrogen doping method

Parasharam M. Shirage[1,†], Kiichi Miyazawa[1,2], Kunihiro Kihou[1,3], Chul-Ho Lee[1,3], Hijiri Kito[1,3], Kazuyasu Tokiwa[2], Yasumoto Tanaka[1], Hiroshi Eisaki[1,3] and Akira Iyo[1,2,3,*]

[1]National Institute of Advanced Industrial Science and Technology, Tsukuba, Ibaraki 305-8568, Japan

[2]Department of Applied Electronics, Tokyo University of Science, Noda, Chiba 275-8510, Japan

[3]JST, Transformative Research-Project on Iron Pnictides (TRIP), 5, Sanbancho, Chiyoda, Tokyo 102-0075, Japan

[†]E-mail: paras-shirage@aist.go.jp

* E-mail: iyo-akira@aist.go.jp

**Abstract:** Here we demonstrate the technique to stabilize the ErFeAsO-based superconductor with the smallest lattice constants in $Ln$FeAsO$_{1-y}$ ($Ln$ = lanthanide) series using hydrogen doping method. Polycrystalline samples were synthesized by heating pellets with nominal compositions of ErFeAsO$_{1-y}$ (1-$y$ = 0.75 ~ 0.95) sandwiched between pellets of LaFeAsO$_{0.8}$H$_{0.8}$ compositions at 1100 °C under a pressure of 5.0 - 5.5 GP. The sample with lattice constants of $a$ = 3.8219 Å and $c$ = 8.2807 Å shows the highest superconducting critical temperatures ($T_c$) of 44.5 K and 41.0 K determined by onset transitions of resistivity and susceptibility, respectively. We discuss phase diagram of $Ln$ dependence of $T_c$ in $Ln$FeAsO-based superconductors.

Superconductivity phase diagrams.

**Introduction.-** The exploration of the iron-based superconductors has created new avenues for the scientific community working in the area of both basic and applied physics [1]. So far, the highest $T_c$ of 55 K has been realized in $Ln$FeAsO-based compounds ($Ln$-1111) crystallized in the ZrCuSiAs-type structure. The superconductivity in $Ln$-1111 is usually induced by transporting electrons from $Ln$O layers to FeAs ones, which can be achieved by various ways like a substitution of $F^-$ for $O^{2-}$ [1], by an oxygen-deficiency [2,3] and by substitution of $Th^{4+}$ for $Ln^{3+}$ [4]. $T_c$ changes with the carrier concentration in the FeAs layers and the highest $T_c$ of each $Ln$-1111 strongly depends on $Ln$. Up to now, $Ln$-1111 samples have been synthesized for $Ln$ = La, Ce, Pr, Nd, Sm, Gd, Tb, Dy, Ho and a rare earth element of Y[5-7]. The lattice constants of $Ln$-1111 contract monotonously with an increase in atomic numbers of $Ln$ i.e. a decrease in ionic radii of $Ln^{3+}$ (lanthanide contraction). The highest $T_c$ increases with the decrease in lattice constants from $Ln$ = La to Nd (Pr) and stays almost constant at around 53 K (or shows a broad peak) between $Ln$ = Nd (Pr) to Dy. Concerning to $Ln^{3+}$ = $Ho^{3+}$ whose ionic radius (1.015 Å) is smaller than $Dy^{3+}$ (1.027 Å) [8], Rodgers *et al.* [7] have reported a suppression of $T_c$ (~ 36 K) in HoFeAs(O,F). On the contrary Yang *et al.* [6] have reported a much higher $T_c$ of 49 K (determined from the onset of diamagnetic transition) in the oxygen-deficient HoFeAsO$_{1-y}$. It should be noted that the highest $T_c$ in $Ln$-1111 usually does not depend on the carrier doping method. The suppression of $T_c$ in HoFeAs(O,F) may be caused by an over-doping effect. Therefore, it is still an open question whether $T_c$ is suppressed for Ho or heavier $Ln$ such as Er, Tm, *etc*. It is important to answer this question to verify the claim that there is a close relationship between $T_c$ and crystal structures such as a bond angle (As-Fe-As) [9-10] and the distance of As atoms from the Fe planes [11-12].

Towards extending the more details on this issue, we have attempted to synthesize ErFeAsO-based superconductors and succeeded to obtain them by applying the hydrogen doping method that we developed to synthesize LaFeAsO$_{1-y}$ samples with shorter lattice constants [13]. In this paper, we report on the synthesis of the Er-1111-based superconductors, its basic physical properties, and we demonstrate a phase diagram of *Ln*-1111 system.

**Experimental Details.-** Polycrystalline samples were synthesized using a cubic-anvil-type HP apparatus and the utility of the HP synthesis technique was demonstrated elsewhere [14, 15]. The most simplest way to dope hydrogen directly into samples, it is desirable to use Er(OH)$_3$ as a starting material. However, Er(OH)$_3$ is not commercially available, so that we used the indirect way (method II) descried in ref. [13] as follows. Starting materials were powders of Fe, Fe$_2$O$_3$ and a precursor of ErAs. The precursor ErAs was synthesized by reacting Er and As chips at 500 °C for 12 h and then followed by heating at 1050 °C for 12 h in an evacuated quartz tube. The starting materials were mixed with a nominal composition of ErFeAsO$_{1-y}$ ($y$ = 0.65 - 0.95), ground and pressed into pellets in a glove box filled with dry nitrogen gas. Besides the two ErFeAsO$_{1-y}$ pellets, we provided three pellets with a nominal composition of LaFeAsO$_{0.8}$H$_{0.8}$ that were made from Fe, LaAs, As and La(OH)$_3$. Five pellets were simultaneously packed into a BN crucible as shown in the inset of Fig. 1, where H-free target ErFeAsO$_{1-y}$ pellets were sandwiched between the LaFeAsO$_{0.8}$H$_{0.8}$ ones. The pellets were heated at a temperature of 1100 °C under a pressure of 5.0 ~ 5.5 GPa for 1 h. The hydrogen ions released from the LaFeAsO$_{0.8}$H$_{0.8}$ pellets were expected to penetrate into the ErFeAsO$_{1-y}$ pellets during sample preparation. NMR measurements suggest that the doped H locates between Fe and La sites [16]. The samples thus synthesized are abbreviated as ErFeAsO$_{1-y}$(H) or Er-1111(H) hereafter. By this technique, we obtained two Er-1111(H) samples with different nominal oxygen contents at the same time. It should be noted that we have never obtained the Er-1111 phase without LaFeAsO$_{0.8}$H$_{0.8}$ pellets.

Powder x-ray diffraction (XRD) patterns were measured using CuK$_\alpha$ radiation

(Ultima IV, Rigaku). The AC magnetic susceptibility was measured under a AC magnetic field amplitude of 0.1 Oe and frequency of 7.7 kHz (PPMS, Quantum Design). The DC magnetic susceptibility was measured using a SQUID magnetometer (MPMS, Quantum Design) under a magnetic field of 5 Oe. The resistivity was measured by a four-probe method.

**Results and Discussion.-**Fig. 1 represents the powder XRD pattern of the (*a*) Er-1111(H) # sample1, (*b*) ErFeAsO$_{1-y}$H (4 GPa) and (*c*) ErFeAsO$_{1-y}$ (5 GPa) as the typical examples. The peaks were indexed based on the ZrCuSiAs-type crystal structure. The Er-1111(H) is a major component along with additional peaks due to impurity phases like ErAs, Er$_2$O$_3$, Fe, Fe$_2$O$_3$, unknown, etc. So far we have synthesized almost singe-phase samples of *Ln*FeAsO$_{1-y}$ for *Ln* = La - Dy by tuning synthesis conditions such as pressure, temperature and oxygen content of starting compositions, etc [5]. However, we have not succeeded in synthesizing a single-phase Er-1111 sample yet even though we had experimented with various synthesis conditions, it resembles with the case of Ho-1111 samples [6,7]. The sample of ErFeAsO$_{1-y}$(H) prepared at a lower pressure of 4 GPa (see Fig. 1(b)) and the sample of ErFeAsO$_{1-y}$ prepared at 5 GPa without the hydrogen doping (see Fig. 1(c)) consist of impurities like ErAs as a main component with/without small trace of the Er-1111 phase. It suggests that synthesis condition is narrow for heavier *Ln* like Er- and Ho-1111. Our previous study demonstrated that a purity of the DyFeAsO$_{1-y}$ sample is improved with increasing synthesis pressure [5]. We may expect to obtain phase pure Er-1111-based samples using higher pressure. However, the pressure limit (~ 5.5 GPa) of our apparatus restricts to test sample synthesis at higher pressure. The lattice constants of typical three samples are listed in Table I, together with unit cell volumes, $T_c$ (definitions of $T_c$ are shown later). The samples with the larger nominal oxygen content have higher $T_c$ values with the bigger lattice constants. The cell volume (121.0 Å$^3$) of the highest $T_c$ sample#3 is smaller than that (121.3 Å$^3$) of HoFeAs(O,F) ($T_c$ = 36.2 K) [7] and that

(122.7 Å$^3$) of HoFeAsO$_{1-y}$ ($T_c$ = 49.1 K)[6].

We extend the discussion here on the Er-1111 phase formation by hydrogen doping into the lattice. The crystal structure of *Ln*FeAsO is constructed by an alternative stacking of *Ln*O and FeAs layers along the *c*-direction. Then, the lattice mismatch along *a*-direction between the *Ln*O and FeAs layers must be small to form the *Ln*FeAsO phase. In the case of *Ln* = La to Gd which can be synthesized even at the ambient pressure, as the lattice mismatch is expected to be small, while it becomes larger for heavier *Ln* such as Ho and Er due to the shrinkage of *Ln*O layers. The lattice mismatch will be reduced under high pressure by a selective compression of the FeAs layers. Consequently, *Ln*FeAsO with heavier *Ln*$^{3+}$ will be formed only under high pressure. Considering the fact that we obtained the Er-1111(H) samples in this study, the doped hydrogen may work to contract FeAs layers.

Fig. 2 shows the temperature (*T*)-dependence of resistivity ($\rho$) for the typical ErFeAsO$_{1-y}$(H) sample#1 - #3. The sample#1 - #3 shows the superconducting transitions ($T_c^{\rho-onset}$) at 35.9, 43.0 and 44.5 K, where ($T_c^{\rho-onset}$) is determined from the intersection of the two extrapolated lines drawn through the resistivity curve just above and below the transition, (see the inset of Fig. 2). The $T_c^{\rho-end}$ was defined as the intersection of the resistive transition curve with the maximum slope on the temperature axis. The resistivity show negative curvatures in the whole measured temperature range above $T_c$. This behaviour is commonly observed for *Ln*FeAsO-based superconductors with heavy *Ln* elements such as Dy and Ho[5,7].

In Fig. 3 *T*-dependence of the in-phase AC susceptibility ($\chi'$) for the sample#1 - #3 are shown. All the samples showed sharp diamagnetic transitions. The superconducting volume fractions estimated by DC magnetic susceptibility from the magnitude of diamagnetic signal (ZFC) at 5 K are about 160 % for the three samples without demagnetizing field corrections ensuring the bulk superconductivity. Onset transition temperature ($T_c^{\chi'-onset}$) is determined from the intersection of the two extrapolated lines drawn through the $\chi'$ curve just above and below the transition (see the inset of Fig. 3). We can clearly define $T_c^{\chi'-onset}$ of 33.2, 39.5 and 41.0 K for the sample#1 - #3, respectively. The $T_c^{\chi'-onset}$ thus determined agrees well with $T_c^{\rho-end}$.

The $a$-lattice constant and the cell volume dependence of $T_c$ for samples including ErFeAsO$_{1-y}$(H) as major phase is shown in Fig. 4. As indicated in Fig. 4 (a), $T_c$ gradually increases with the cell volume. All the samples synthesized under a pressure of 5.0 GPa showed larger lattice constants and higher $T_c$ ($\geq$ 40 K), while the other ones synthesized in higher pressure of 5.5 GPa showed lower $T_c$ (< 40 K) as indicated in Fig. 4. One may expect that samples prepared under lower pressure (less than 5.0 GPa) will make $T_c$ higher. However, it is an experimentally difficult task to synthesize samples including Er-1111 as a main component at a lower pressure, most probably as the pressure is not enough to reduce the lattice mismatch between ErO and FeAs layers as discussed above. On the other hand, $T_c$ appears to have a broad peak around $a = 3.822$ Å, as one can see in Fig. 4 (b). Because it is difficult to change the $a$- and $c$-lattice constants independently so that a large increase of $T_c$ may not be expected in Er-1111(H) even though samples can be synthesized under lower pressure.

The highest $T_c$ (determined from onsets of diamagnetic transitions) obtained so far in our group are plotted against the $a$-lattice constant in Fig. 5 for $Ln$FeAsO$_{1-y}$ ($Ln$ = La, (La,Y), Ce, Pr, Nd, Sm, Gd, Tb and Dy) [5,17,18] and $Ln$FeAsO$_{1-y}$(H) ($Ln$ = La, Ce, Pr and Sm) [13,19] together with HoFeAs(O,F) [7], HoFeAsO$_{1-y}$ [6] and Er-1111(H) superconductors. The $T_c$ of the HoFeAs(O,F) is lower than that of the Er-1111(H) though it has longer $a$-lattice constant and larger cell volume. Rodgers *et al*.[7] reported that the positive $dT_c/dV$ observed in HoFeAs(O,F) as a particular property of HoFeAs(O,F). However, the positive $dT_c/dV$ is also observed in the over-doped region of $Ln$FeAsO$_{1-y}$ ($Ln$ = La[18], Nd[3] and Sm[19]) and LaFeAs(O,F) [20]. In addition, we have confirmed that the highest $T_c$ (49 K) of HoFeAsO$_{1-y}$ agrees with that reported by Yang et al. [6] by constructing the phase diagram [21]. This evidence suggested that suppression of $T_c$ in HoFeAs(O,F) may be due to over-doping effect rather than intrinsic suppression due to the lattice contraction. Similarly, one may have suspicions that lower $T_c$ in ErFeAsO$_{1-y}$(H) is due to disorder caused by the hydrogen doping into the samples. As shown in Fig. 5, $T_c$ of $Ln$FeAsO$_{1-y}$(H) ($Ln$ = La, Ce and Pr) moves almost on the curve which is drawn by $Ln$FeAsO$_{1-y}$ samples so that the hydrogen doping mainly affects the lattice contraction and $T_c$ changes, consequently. Now we can draw a clear peak at about $Ln$ = Sm in the *a*-axis dependence of $T_c$ as indicated in Fig. 5. This figure demonstrates that suitable lattice size for higher $T_c$ exists in $Ln$-1111 superconductors.

We have succeeded to stabilize the heavy lanthanide Er-1111-based superconductors for the first time by the hydrogen doping method. The sample with lattice constants of *a* = 3.8219 Å and *c* = 8.2807 Å shows the highest $T_c^{\chi'-onset}$ of 41.0 K. We demonstrated that the Er-1111-based superconductors having the smallest lattice constants in $Ln$FeAsO$_{1-y}$ series have much lower $T_c$ than HoFeAsO-based one. We have established the phase diagram of the *a*-lattice constant *vs*. $T_c$ for $Ln$-1111-based superconductors from $Ln$ = La up to Er, where a clear peak at about $Ln$ = Sm can be seen.

**Acknowledgements:**

This work was supported by a Grant-in-Aid for Specially Promoted Research (20001004) from The Ministry of Education, Culture, Sports, Science and Technology (MEXT) and JST, Transformative Research-Project on Iron Pnictides (TRIP).

**References:**

[1] Kamihara Y., Watanabe T., Hirano M., and Hosono H., J. Am. Chem. Soc. **130**, (2008) 3296.

[2] Kito H., Eisaki H., and Iyo A., J. Phys. Soc. Jpn. **77**, (2008) 063707.

[3] Ren Z. A., Che G. C., Dong X. L., Yang J., Lu W., Yi W., Shen X. L., Li Z. C., Sun L. L., Zhou F., and Zhao Z. X., Europhys. Lett. **83,** (2008) 17002.

[4] Wang C., Li L., Chi S., Zhu Z., Ren Z., Li Y., Wang Y., Lin X., Luo Y., Jiang S., Xu X., Cao G., and Xu Z., Europhys. Lett. **83**, (2008)67006.

[5] Miyazawa K., Kihou K., Shirage P. M., Lee C. H., Kito H., Eisaki H., and Iyo A., J. Phys. Soc. Jpn. **78**, (2009)034712.

[6] Yang J., Shen X.L., Lu W., Yi W., Li Z. C., Ren Z. A., Che G. C., Dong X. L., Sun L. L., Zhou F., and Zhao Z. X., New J. Physics **11,** (2009) 0255005.

[7] Rodgers J. A., Penny G. B. S., Marcinkova A., Bos J.-W. G., Sokolov D. A., Kusmartseva A., Huxley A. D., and Attfield J. P., Phys. Rev. B **80**, (2009) 052508.

[8] Shannon R. D.: Acta Cryst. A **32,** (1976) 751.

[9] Lee C.H., Iyo A., Eisaki H., Kito H., Fernandez-Diaz M. T., Ito T., Kihou K., Matsuhata H., Braden M., and Yamada K., J. Phys. Soc. Jpn. **77,** (2008) 083704.

[10] Saito T., Onari S., and Kontani H., Phys. Rev. B **82**, (2010)144510.

[11] Mizuguchi Y., Hara Y., Deguchi K., Tsuda S., Yamaguchi T., Takeda K., Kotegawa H., Tou H., and Takano Y., Supercond. Sci. Technol. **23,** (2010)54013.

[12] Kuroki K., Usui H., Onari S., Arita R., and Aoki H., Phys. Rev. B **79**, (2009)224511; Sawatzky G. A., Elfimov I. S., Brink J. van den, and Zaanen J., Europhys. Lett. **86**, (2009)17006.

[13] Miyazawa K., Ishida S., Kihou K., Shirage P. M., Nakajima M., Lee C. H., Kito H., Tomioka Y., Ito T., Eisaki H., Yamashita H., Mukuda H., Tokiwa K., Uchida S., and Iyo A., App. Phys. Lett. **96**, (2010)072514.

[14] Shirage P. M., Miyazawa K., Ishikado M., Kihou K., Lee C.H., Takeshita N., Matsuhata H., Kumai R., Tomioka Y., Ito T., Kito H., Eisaki H., Shamoto S., and Iyo A., Physica C **469**, (2009) 355.

[15] Shirage P. M., Kihou K., Miyazawa K., Lee C. H., Kito H., Eisaki H., Yanagisawa T., Tanaka Y., and Iyo A., Phys. Rev. Lett. **103,** (2009) 257003.

[16] Yamashita H., Mukuda H., Yashima M., Furukawa S., Kitaoka Y., Miyazawa K., Shirage P. M., Eisaki H., and Iyo A., J. Phys. Soc. Jpn. **79,** (2010) 103703.

[17] Shirage P. M., Miyazawa K., Kito H., Eisaki H., and Iyo A., Phys. Rev. B **78,** (2008) 172503.

[18] Miyazawa K., Kihou K., Ishikado M., Shirage P. M., Lee C. H., Takeshita N., Eisaki H., Kito H., and Iyo A., New J. Phys. **11,** (2009)45002.

[19] Shirage P. M., Kihou K., Miyazawa K., Kito H., Yoshida Y., Tanaka Y., Eisaki H., and Iyo A., Phys. Rev. Lett. **105,** (2010) 037004.

[20] Wakimoto S., Kodama K., Ishikado M., Matsuda M., Kajimoto R., Arai M., Kakurai K., Esaka F., Iyo A., Kito H., Eisaki H., and Shamoto S., J. Phys. Soc. Jpn. **79,** (2010) 74715.

[21] Shirage P. M. and Iyo A., Unpublished data.

**Figure captions:**

**Fig. 1** Powder XRD patterns of (a) Er-1111(H) prepared at 5.0 GPa (sample#1) as a typical example along with (b) Er-1111(H) prepared at 4 GPa and (c) Er-1111 prepared at 5 GPa without the hydrogen doping. Peaks are indexed on the basis of the tetragonal structure (P4/nmm). Inset shows the cross section of sample cell assembly for high-pressure synthesis.

**Fig. 2** Temperature dependence of resistivity ($\rho$) for ErFeAsO$_{1-y}$(H) sample#1 - #3. The inset shows the resistivity near superconducting transition. Definitions of $T_c^{\rho-onset}$ and $T_c^{\rho-end}$ are indicated in the inset.

**Fig. 3** Temperature dependence of AC susceptibility ($\chi'$) for ErFeAsO$_{1-y}$(H) sample#1 - #3. The inset shows the $\chi'$ near superconducting transition. A definition of $T_c^{\chi'-onset}$ is indicated in the inset.

**Fig. 4 (a-b)** Cell volume (a) and $a$-lattice constant (b) dependence of $T_c$ for samples which include ErFeAsO$_{1-y}$(H) phase as a major component. The samples were synthesized under a pressure of 5.0 GPa (closed circles) or 5.5 GPa (open circles). $T_c$'s are determined from onset of diamagnetic transitions from DC magnetization. The dashed line and curve are guides to the eye.

**Fig. 5** Variation of $T_c$ determined from the onset of diamagnetic transitions for $Ln$FeAsO-based superconductors. The highest $T_c$ obtained so far in our group are plotted for $Ln$FeAsO$_{1-y}$ ($Ln$ = La, (La,Y), Ce, Pr, Nd, Sm, Gd, Tb and Dy) [5,17] and $Ln$FeAsO$_{1-y}$(H) ($Ln$ = La, Ce, Pr and Sm) [13,19] together with HoFeAs(O,F) [7], HoFeAsO$_{1-y}$ [6] and Er-1111(H) superconductors.

Table I: Lattice constants of typical three samples are listed along with the unit cell volumes and $T_c$. The sample#1, #2 and #3 were synthesized under pressures of 5.5, 5.0 and 5.0 GPa, respectively.

| Samples | Nominal starting compositions | $a$ (Å) | $c$ (Å) | Vol. (Å$^3$) | $T_c^{\rho-onset}$ (K) | $T_c^{\rho-end}$ (K) | $T_c^{\chi'-onset}$ (K) |
|---|---|---|---|---|---|---|---|
| sample#1 | ErFeAsO$_{0.75}$ | 3.8198 | 8.2517 | 120.40 | 35.9 | 33.5 | 33.2 |
| sample#2 | ErFeAsO$_{0.95}$ | 3.8238 | 8.2680 | 120.89 | 43.0 | 40.6 | 39.5 |
| sample#3 | ErFeAsO$_{0.95}$ | 3.8219 | 8.2807 | 120.96 | 44.5 | 41.9 | 41.0 |

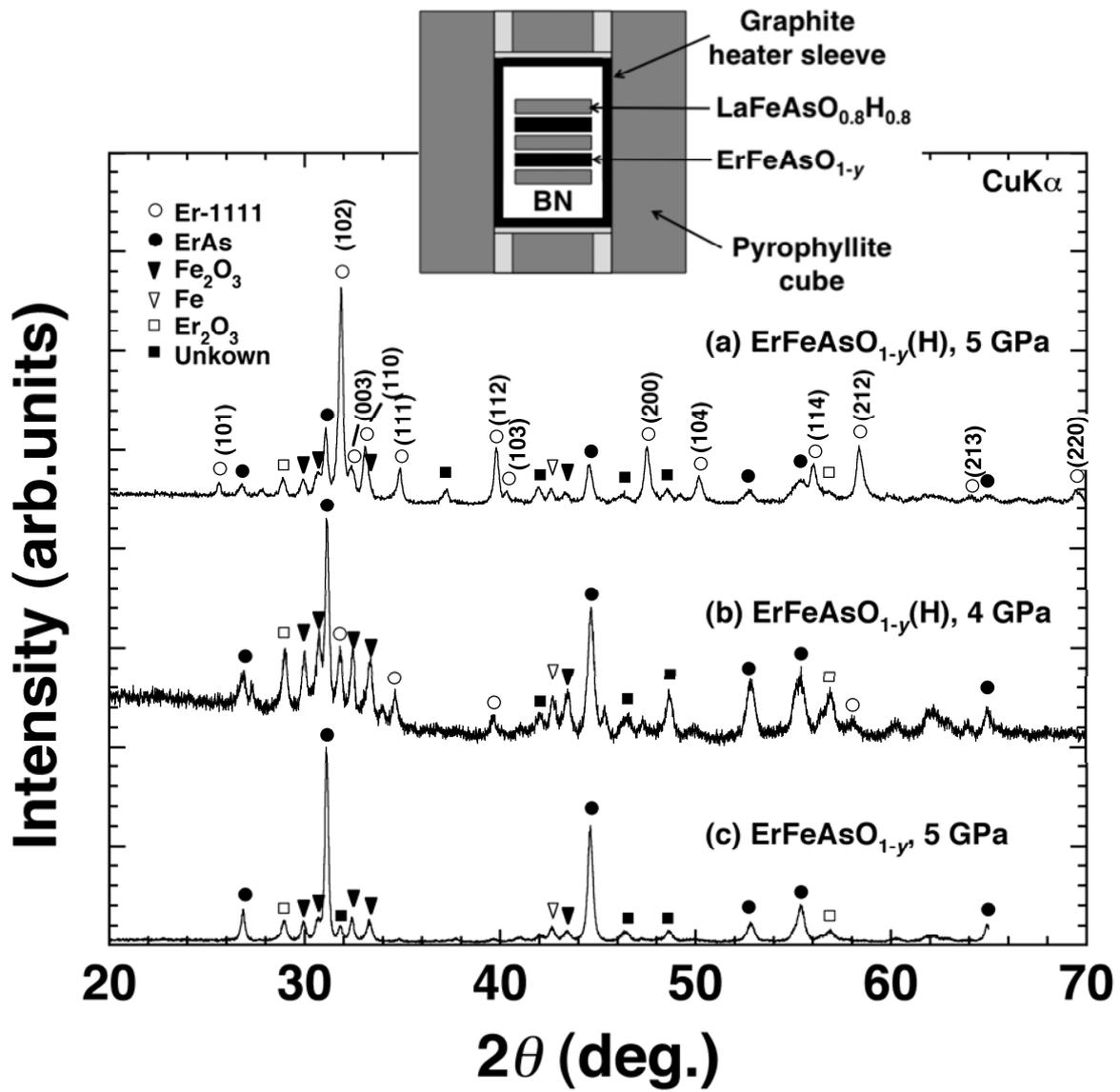

**Figure 1-Shirage-**

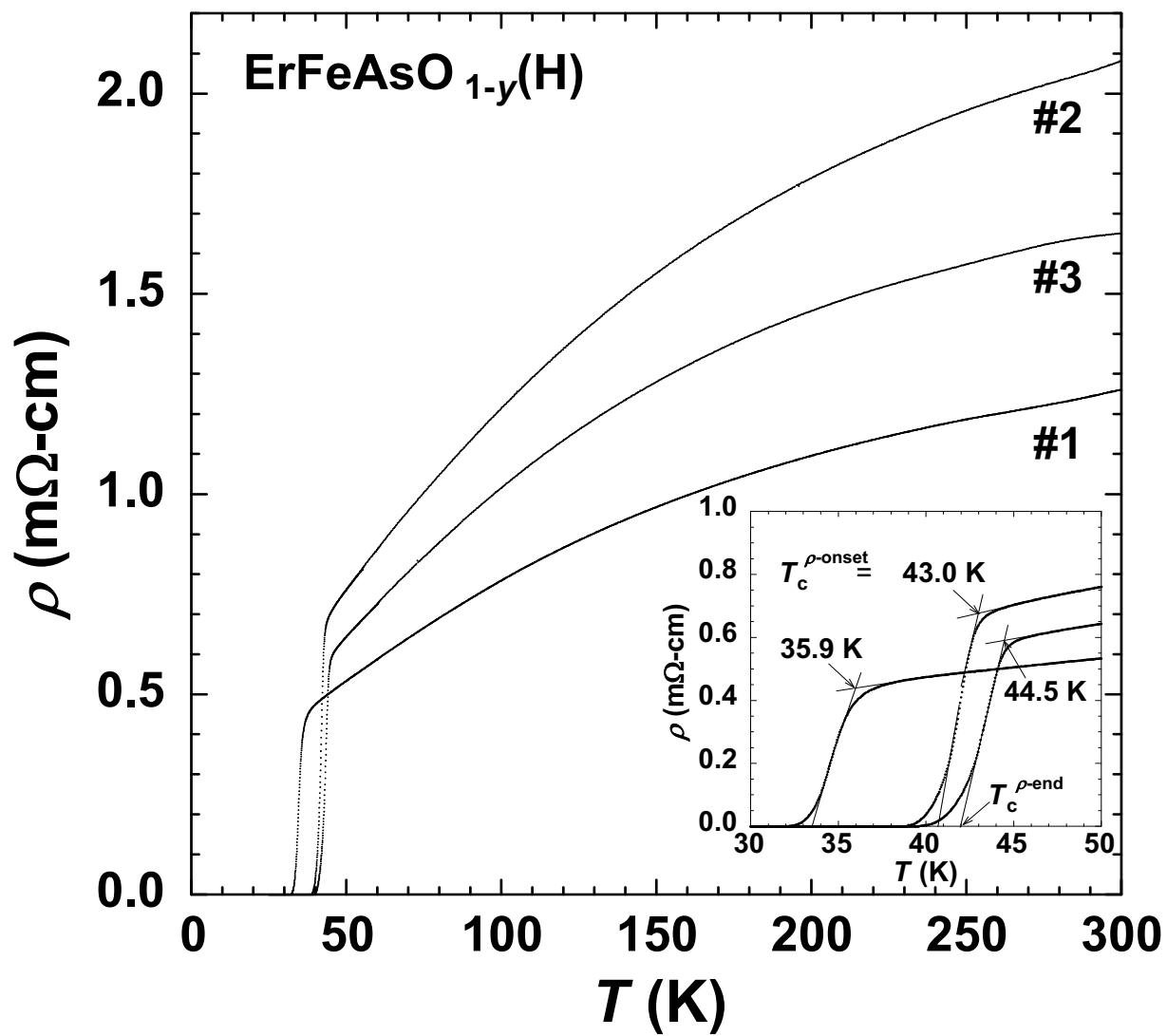

Figure 2-Shirage-

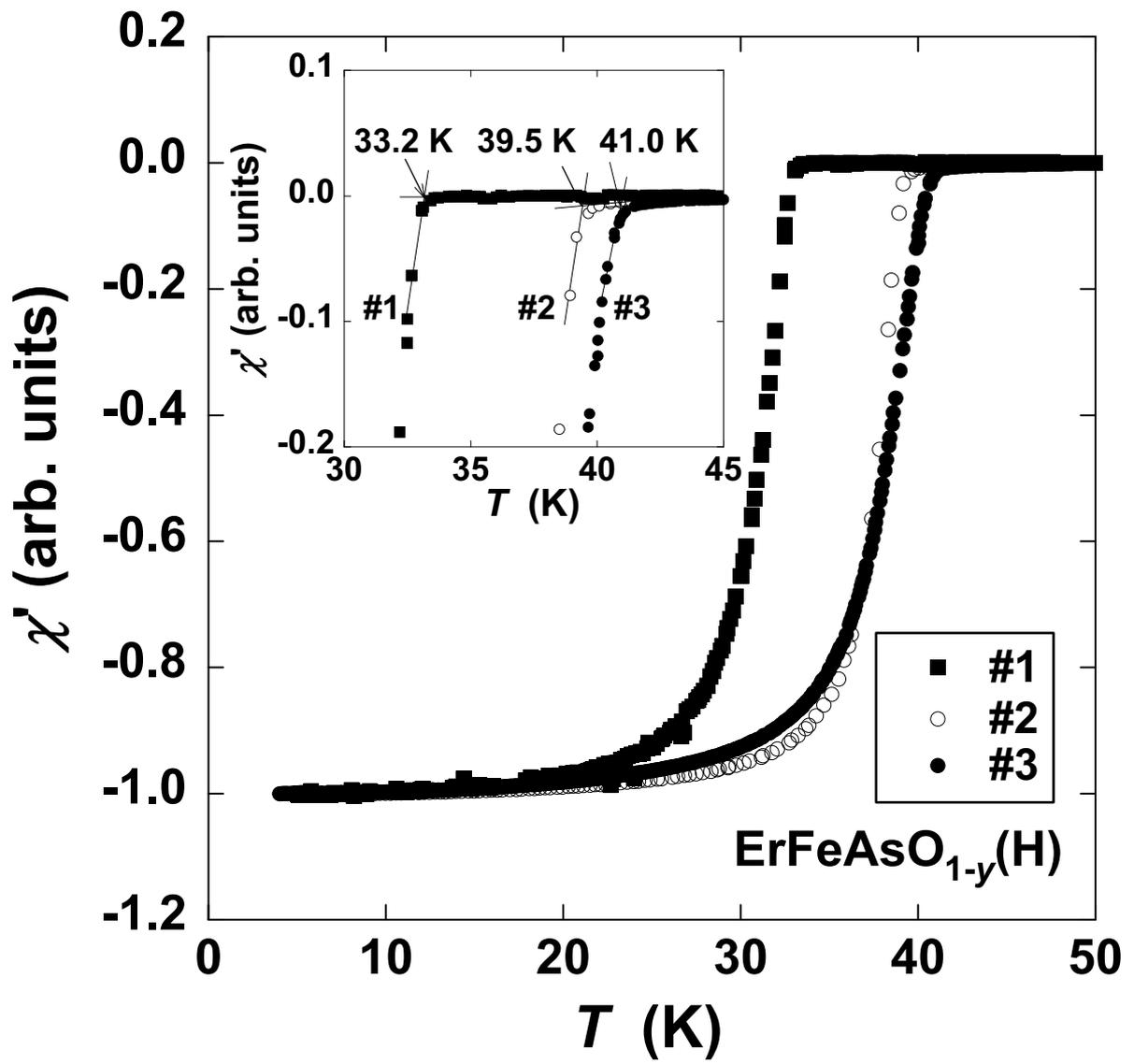

**Figure 3-Shirage-**

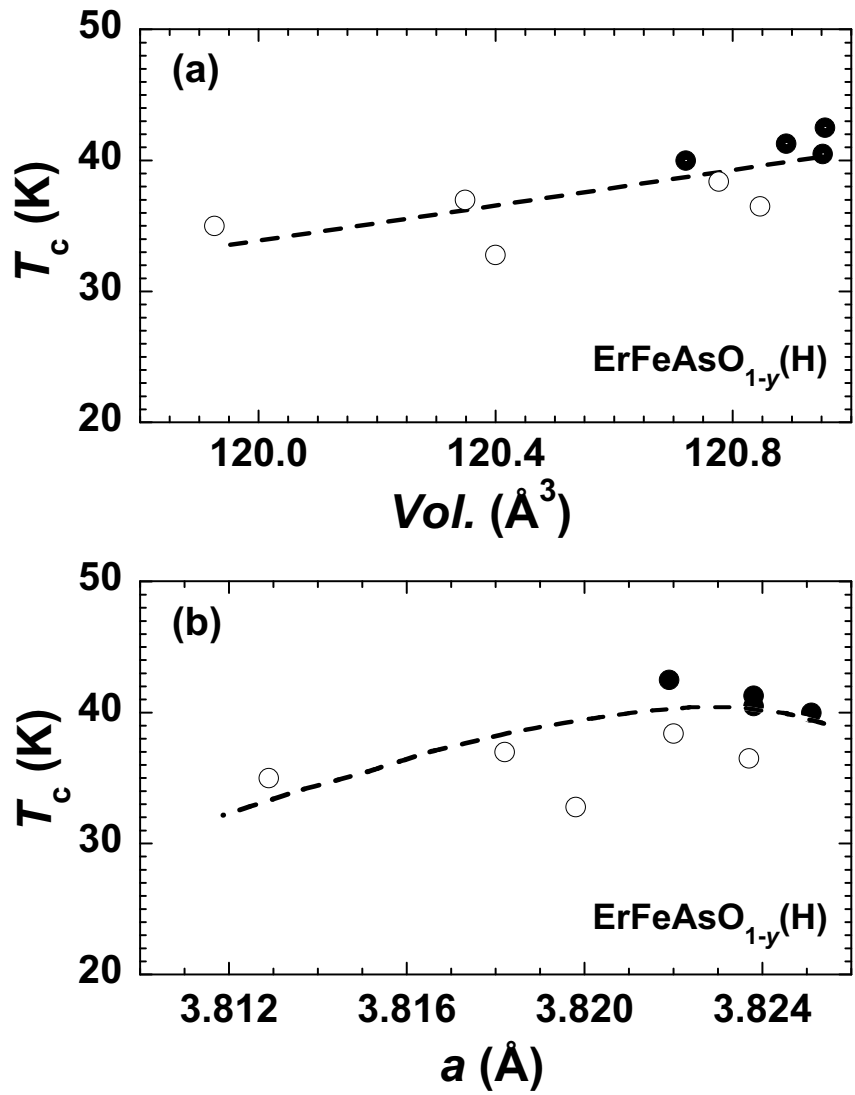

**Figure 4-Shirage-**

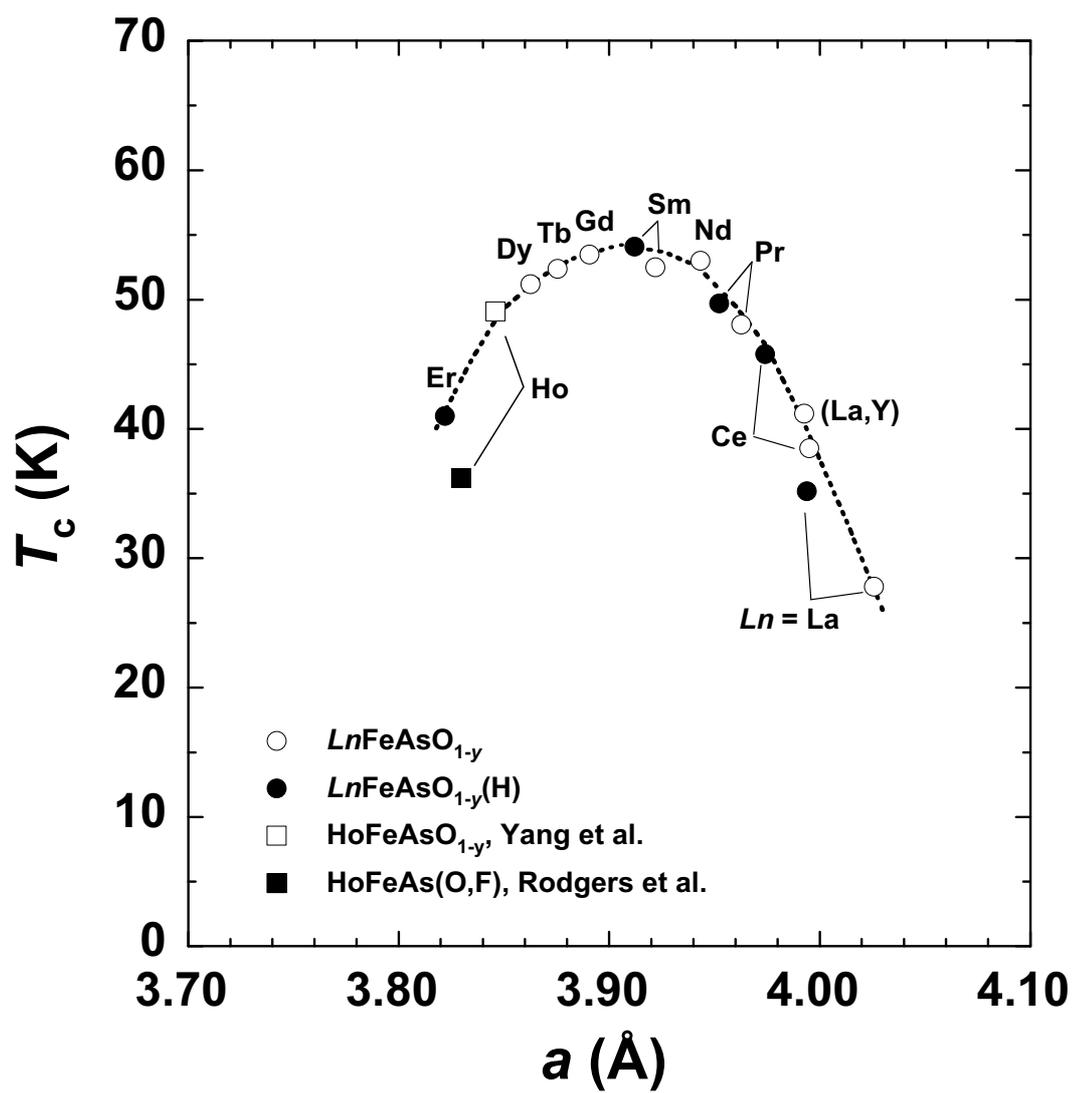

**Figure 5-Shirage-**